\documentclass[
  aps,
  prx,
  floatfix,
  superscriptaddress,
  longbibliography,
  reprint
]{revtex4-2}

\usepackage[utf8]{inputenc}
\usepackage{mathtools}
\usepackage{physics}
\usepackage{graphicx}
\usepackage{subfigure}
\usepackage{array}
\usepackage{amsmath}
\usepackage[colorlinks=true,allcolors=blue]{hyperref}
\usepackage[capitalize]{cleveref}

\begin{document}

\title{Quantum Simulation of Dissipative Energy Transfer via Noisy
QuantumComputer}

\author{Chin-Yi Lin}
\affiliation{Department of Physics, National Taiwan University, Taipei 10617, Taiwan}

\author{Shin Sun}
\affiliation{Quantum Information Science, Okinawa Institute of Science and Technology, Okinawa 904-0495, Japan}

\author{Li-Chai Shih}
\affiliation{Department of Chemistry, National Taiwan University, Taipei 10617, Taiwan}

\author{Yuan-Chung Cheng}
\affiliation{Department of Chemistry, National Taiwan University, Taipei 10617, Taiwan}

\date{June 2023}
\begin{abstract}
We study whether dissipative energy-transfer dynamics can be simulated on noisy near-term quantum hardware by treating device noise as a calibrated resource rather than purely as an error source. Focusing on a biased exciton dimer, we encode the single-excitation manifold into a two-qubit subspace and implement the coherent dynamics through a shallow Trotterized propagator, while repeated noisy identity operations provide an effective dissipative channel. We benchmark the resulting short-time population dynamics against the hierarchical equations of motion (HEOM), which serves as a numerically accurate reference for the corresponding open-system model. On IBM quantum hardware, the calibrated noisy circuit reproduces a broad range of dissipative trajectories in the tested regime, and the fitted HEOM parameters exhibit an approximately linear dependence on the noisy-gate frequency. This empirical relation enables a practically useful interpolation strategy: once calibrated by a finite set of HEOM calculations, the noisy circuit can replace repeated HEOM fitting for intermediate parameter points within the same biased-dimer family. To extend the dynamics beyond the circuit-depth limit, we combine the short-time quantum data with the transfer tensor method (TTM). In simulator studies, TTM accurately extends the dynamics well beyond the directly simulated window, whereas on real hardware its performance is limited by the instability of coherence-sensitive initial states. Our results show that noisy few-qubit devices can act as calibrated phenomenological simulators of open-system dynamics and, within a restricted but experimentally relevant regime, can provide a practical surrogate for repeated HEOM-based modeling.
\end{abstract}

\maketitle

\section{Introduction}
Quantum simulation is one of the most compelling applications of quantum computing because it offers a direct route to studying many-body and dynamical quantum phenomena that are difficult to compute classically. In practice, however, current quantum hardware remains in the noisy intermediate-scale quantum (NISQ) regime, where limited qubit counts and imperfect gate operations strongly constrain the complexity of executable algorithms. As a result, near-term quantum simulation methods must be designed not only around the target physics but also around the noise properties of the hardware itself.

Open quantum systems provide a particularly important and challenging setting for this effort. Because the system interacts with an environment, its dynamics generally exhibit decoherence, dissipation, and memory effects that are absent in isolated systems. These effects are central to problems in chemical physics, condensed-matter physics, and quantum information science. For excitation energy transfer problems, accurate simulation often requires going beyond weak-coupling or Markovian approximations. Numerically accurate classical approaches such as the hierarchical equations of motion (HEOM) can provide reliable benchmarks, but their computational cost grows rapidly with model complexity and bath-memory requirements.

Recent work by Sun et al.~\cite{sun2021efficient} showed that gate noise on present-day quantum hardware can be used constructively to emulate dissipative dynamics in a symmetric exciton dimer. That perspective is especially appealing in the NISQ setting because it treats noise not only as a limitation, but also as a potential computational resource. The present work extends that idea to a more general and less symmetric model: a biased exciton dimer with unequal site energies. In this case, the effective system Hamiltonian contains both $\sigma_Z$ and $\sigma_X$ terms, so the coherent propagator is more difficult to implement than in the symmetric setting. To address this, we construct the unitary part of the evolution through a Trotterized decomposition that can be executed with shallow two-qubit circuits.

The central question of this study is whether noisy two-qubit quantum hardware can reproduce physically meaningful dissipative energy-transfer dynamics in this more general biased-dimer model. Our approach combines two ingredients. First, we encode the two-site system into a two-qubit subspace and use repeated noisy identity operations to emulate the dissipative contribution of the environment. Second, we benchmark the resulting short-time dynamics against HEOM in order to determine how the hardware-induced noise maps onto effective open-system parameters. This comparison reveals an approximately linear relation between the noisy-gate frequency and the fitted HEOM parameters over the tested regime, providing a practical calibration strategy for short-time simulations.

Long-time simulation remains difficult because the Trotterized propagator requires deeper circuits as the evolution time increases, making the results progressively more sensitive to hardware noise. To mitigate this limitation, we combine the short-time quantum trajectories with the transfer tensor method (TTM), which reconstructs long-time non-Markovian dynamics from a finite training window. In this work, TTM performs well when applied to simulator data, while its performance on real-device data is limited by the instability of coherence-sensitive initial states.

This paper is organized as follows. We first introduce the biased exciton-dimer model and the HEOM benchmark for dissipative dynamics. We then describe the two-qubit encoding, the Trotterized implementation of the system propagator, and the noisy circuit element used to emulate dissipation. Next, we compare the resulting short-time quantum dynamics with HEOM and discuss the empirical relation between hardware noise and effective dissipative parameters. Finally, we apply TTM to extend the short-time dynamics and assess its validity on both simulator and real-device data.

\begin{figure*}[t]
\centering
\subfigure[Quantum simulation]{\includegraphics[width=0.5\textwidth,height=0.2\textwidth]{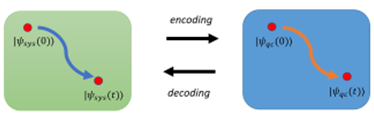}}
\subfigure[Energy transfer in a biased two-site system]{\includegraphics[width=0.4\textwidth,height=0.2\textwidth]{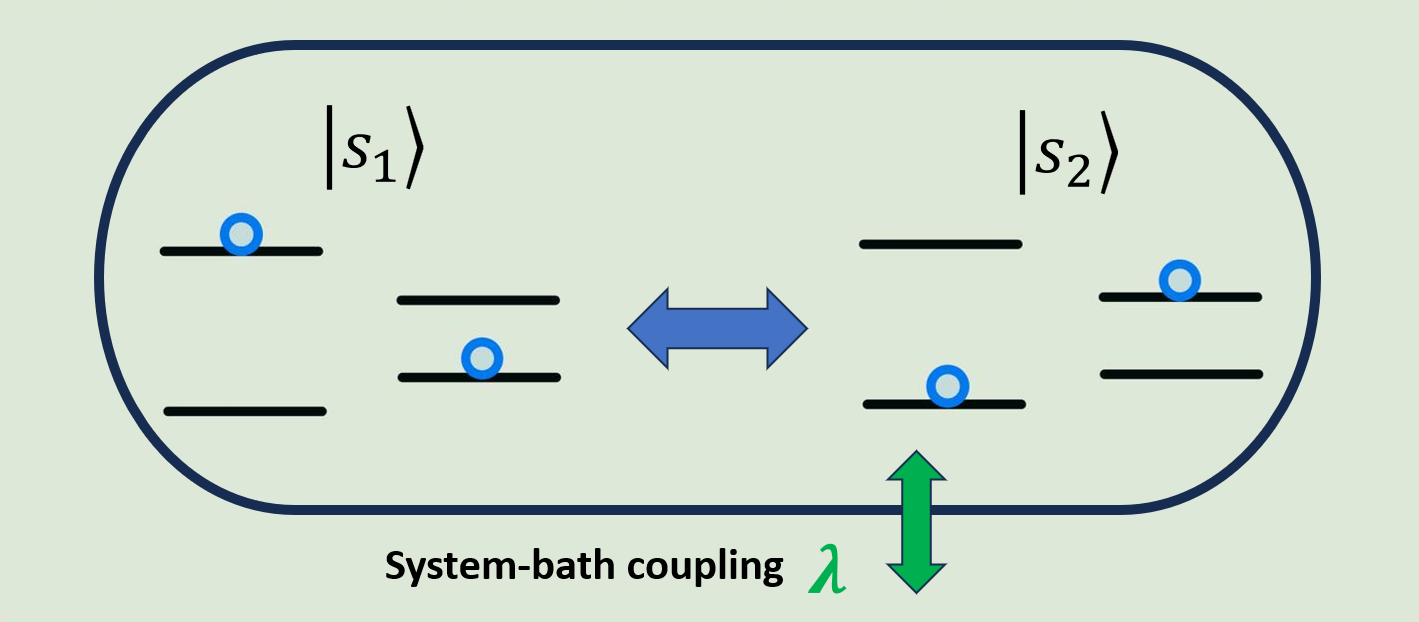}}
\caption{Schematic overview of the problem studied in this work. A noisy quantum circuit is used to simulate dissipative energy transfer in a biased two-site system.}
\label{Fig.main2}
\end{figure*}

\section{Energy Transfer in Open Quantum System}

\subsection{Bias Exciton Dimer System}
We consider excitation energy transfer in a two-site exciton dimer. Each site is treated as a two-level subsystem, and the effective system Hamiltonian is written in second-quantized form as
\begin{equation}\label{Eq.Hamiltonian}
H=\sum_{i=1}^{2}\epsilon_{i}a^{\dagger}_{i}a_{i}+\sum_{i\neq j=1}^{2} J_{ij}a_{i}^{\dagger}a_{j},
\end{equation}
where $a^{\dagger}_{i}$ and $a_i$ are the creation and annihilation operators for site $i$, $\epsilon_i$ denotes the site energy, and $J_{ij}$ denotes the coupling between the localized states $\ket{i}$ and $\ket{j}$.

Unlike the symmetric dimer studied in Ref.~\cite{sun2021efficient}, we focus on the biased case $\epsilon_{1}\neq\epsilon_{2}$. This extension is physically relevant because realistic energy-transfer systems generally do not possess exact site symmetry. In the single-excitation manifold, the dynamics are governed by the site-energy bias and the inter-site coupling, which together define an effective two-level problem that is more general than the previously studied symmetric setting.

\subsection{Open Quantum System}
If the dimer were isolated, its dynamics would be governed by coherent population transfer between the two sites. In realistic environments, however, the system is coupled to surrounding degrees of freedom, and this coupling induces decoherence and dissipation. The resulting dynamics therefore belong to the broader class of open quantum systems. Since our goal is to model excitation transfer under physically relevant conditions, the environmental contribution cannot be neglected. A schematic representation of the target problem is shown in \cref{Fig.main2}.

\subsection{Hierarchical Equations of Motion(HEOM)}
As a classical benchmark, we use the hierarchical equations of motion (HEOM) to compute the dissipative dynamics of the dimer. In this framework, the system-bath interaction is described by the Drude-Lorentz spectral density
\begin{equation}\label{Eq. Drude-Lorentz}
J(\omega)=\frac{\lambda}{2}\frac{\gamma\omega}{\gamma^{2}+\omega^{2}},
\end{equation}
where $\lambda$ is the reorganization energy characterizing the system-bath coupling strength, and $\gamma$ is the cutoff frequency associated with the bath relaxation timescale. This model provides a standard description of the dissipative environment relevant to excitation energy transfer.

HEOM is widely recognized as a reliable method for non-Markovian open-system dynamics, and in this work it serves as our reference calculation. At the same time, accurate HEOM simulations can become computationally demanding because they require hierarchical truncation and the propagation of a large set of coupled equations. We therefore use HEOM as a benchmark against which the quantum-hardware results can be calibrated and assessed.

\section{Quantum Simulation Using Real Device Quantum Computer}

\begin{figure*}[t]
\centering
\subfigure[Circuit overview]{\includegraphics[width=0.4\textwidth]{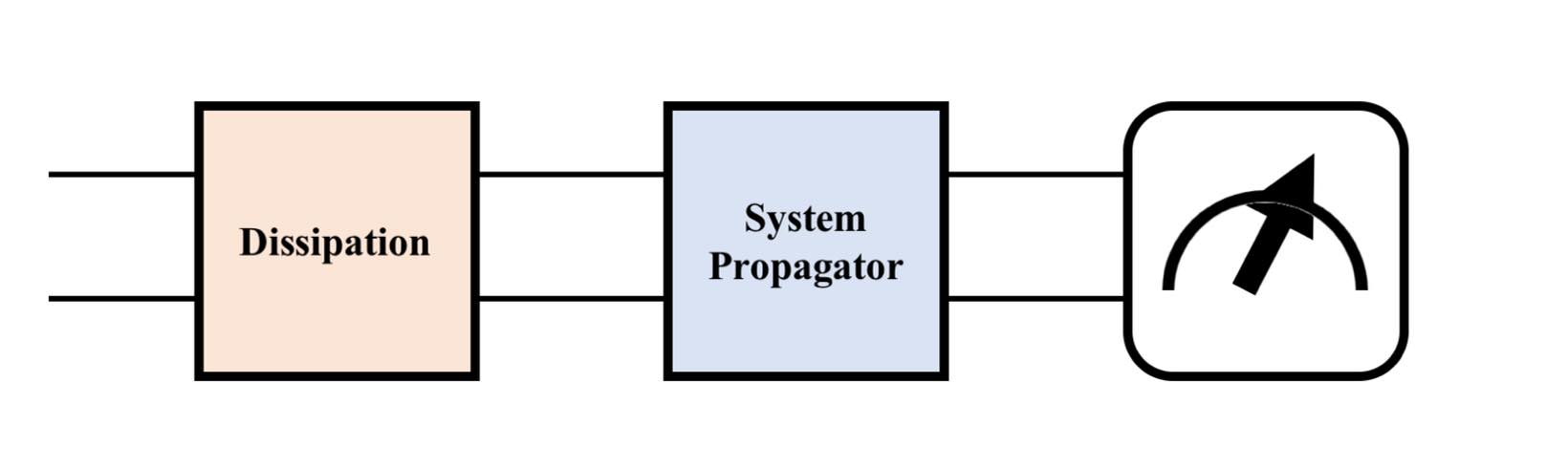}}
\subfigure[A Trotter step of the system propagator]{\includegraphics[width=0.5\textwidth]{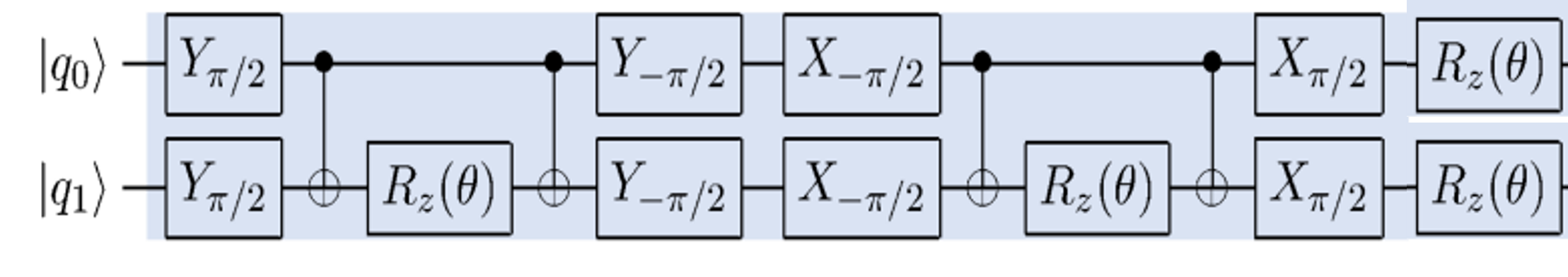}}
\subfigure[Dissipative circuit]{\includegraphics[width=0.5\textwidth]{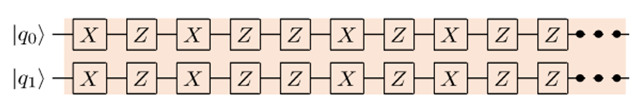}}
\caption{Structure of the quantum circuit. The dissipative component is implemented through repeated noisy identity gates of the form $(XZXZZ)^{2\delta_Q t}$, while the coherent component implements the effective system propagator $\exp(-iHt)$.}
\label{Fig.Scheme}
\end{figure*}

Our quantum circuit consists of two conceptually distinct components: a unitary propagator that reproduces the coherent part of the dimer dynamics, and a noisy gate sequence that provides an effective dissipative channel. The overall strategy is to keep the circuit small enough for present-day hardware while using device noise in a controlled and interpretable way. The construction is described below.

\subsection{State Encoding}
To describe excitation transfer between the two sites, we restrict attention to the single-excitation subspace spanned by $\ket{s_1}$ and $\ket{s_2}$. Here $\ket{s_1}$ denotes the configuration in which site 1 is excited and site 2 is in the ground state, while $\ket{s_2}$ denotes the opposite configuration. Without loss of generality, we label $\ket{s_1}$ as the higher-energy state. States with zero or double excitation are neglected because the present model focuses on single-excitation dynamics.

We encode these two basis states into two qubits by identifying $\ket{10}$ with $\ket{s_1}$ and $\ket{01}$ with $\ket{s_2}$. Within this encoding, population transfer between the two computational states directly represents excitation transfer in the dimer. For a quantum state $\ket{\psi}$ generated by the two-qubit circuit, and neglecting leakage into $\ket{00}$ and $\ket{11}$, we have
\begin{equation}\label{Eq. quantum state mapping}
\ket{\psi}=c_1\ket{s_1}+c_2\ket{s_2}=c_1\ket{10}+c_2\ket{01}.
\end{equation}
Repeated measurement yields the populations $p_1=\abs{c_1}^2$ and $p_2=\abs{c_2}^2$, which correspond to the site populations in the dimer and, equivalently, to the diagonal elements of the reduced density matrix $\rho_S(t)$ in the site basis. Throughout this work, we focus primarily on the dynamics of $p_1(t)$, the population of the higher-energy site.

\subsection{System Propagator}
\subsubsection{System Hamiltonian}
The coherent part of the simulation is governed by the Schr\"odinger equation. Setting $\hbar=1$, the formal time evolution is
\begin{equation}
\ket{\Psi}=e^{-iHt}\ket{\Psi_{0}} ,
\end{equation}
where the Hamiltonian is given in \cref{Eq.Hamiltonian}. In the single-excitation manifold, the diagonal contribution represents the site-energy bias and the off-diagonal contribution represents tunneling between the two sites. Since the identity component contributes only a global phase, the physically relevant parameters are the bias $2\epsilon=\epsilon_{1}-\epsilon_{2}$ and the coupling $J$.

For the biased dimer considered here, the effective Hamiltonian can be written in matrix form as
\begin{equation}
H=\begin{bmatrix} \epsilon&J\\ J&-\epsilon \end{bmatrix}=(\epsilon \sigma_Z+ J\sigma_X).
\end{equation}
Compared with the symmetric case, the simultaneous presence of $\sigma_Z$ and $\sigma_X$ terms makes the implementation of the propagator less straightforward on quantum hardware.

\subsubsection{Quantum Circuits of the System Propagator}
Because $[\sigma_X,\sigma_Z]\neq0$, the propagator $e^{-iHt}$ cannot be factorized exactly into independent $\sigma_X$ and $\sigma_Z$ rotations. We therefore approximate the evolution by a first-order Trotter decomposition:
\begin{equation}\label{expansion}
\begin{split}
e^{-iHt}
&=e^{-i(\epsilon \sigma_Z+ J \sigma_X)t} \\
&= e^{-itJ \sigma_X}e^{-it\epsilon \sigma_Z}
e^{\frac{-t^2\epsilon J}{2} [ \sigma_Z, \sigma_X]}e^{O(t^{3})}
\end{split}
\end{equation}
For sufficiently small $\delta t$, the higher-order commutator term becomes negligible, giving
\begin{equation}\label{Eq. small t approximation}
\begin{split}
e^{-iH\delta t}
&= e^{-i\delta tJ \sigma_X}e^{-i\delta t\epsilon \sigma_Z}
e^{\frac{-\delta t^2\epsilon J}{2} [ \sigma_Z, \sigma_X]}e^{O(\delta t^{3})} \\
&\approx e^{-i\delta tJ \sigma_X}e^{-i\delta t\epsilon \sigma_Z}.
\end{split}
\end{equation}
This approximation allows each short-time propagation step to be implemented using elementary rotations associated with the $\sigma_X$ and $\sigma_Z$ components.

For a longer evolution time, we divide the interval into $M$ Trotter steps, $t=M\Delta t$, with $\Delta t\ll1$:
\begin{equation}\label{Trotter Split}
\begin{split}
e^{-iHt}
&=\prod e^{-iH\Delta t}
=\prod e^{-i(\epsilon \sigma_Z+ J \sigma_X)\Delta t} \\
&\approx \prod e^{-i\Delta tJ \sigma_X}e^{-i\Delta t\epsilon \sigma_Z}
\end{split}
\end{equation}
Increasing $M$ improves the Trotter approximation by keeping each time slice small. In practice, we choose $M$ to scale with the simulated evolution time so that $\Delta t=t/M$ remains in a regime where the approximation is reliable. The corresponding circuit structure is illustrated in \cref{Fig.Scheme}.

The tradeoff is that the circuit depth grows with $M$, and hence with the simulated time. On a real device, this means that long-time propagation suffers increasingly from gate errors. This depth-noise tradeoff is one of the main reasons we later introduce TTM to extend the dynamics without continuing to deepen the circuit.

\subsection{Dissipative Part of the Quantum Circuit}
The dissipative part of the dynamics is not implemented through an explicit bath register. Instead, we use the intrinsic gate noise of the hardware as an effective source of decoherence. The purpose of this construction is not to claim microscopic equivalence between device noise and a physical bath, but to test whether calibrated noisy circuits can reproduce the phenomenology of dissipative energy transfer.

\subsubsection{Dissipation Circuits}
In our study, this effective noise is injected by repeatedly applying the identity sequence $I=(XZXZZ)^{2}$. Further discussion of this gate choice is provided in the supplementary information. To emulate different values of the HEOM system-bath coupling strength $\lambda$, we apply a number of noisy identity gates that grows linearly with time. For the data point at time $t$, the number of such operations is $N=\delta_Q t$, where $\delta_Q$ denotes the frequency of the noisy identity gate. The dissipation circuit is shown in \cref{Fig.Scheme}. A larger $\delta_Q$ leads to faster decoherence in the measured dynamics. By varying $\delta_Q$, we generate a family of trajectories with different effective damping strengths that can be compared with HEOM results.

\subsection{Post Processing}
After the dissipative sequence and the coherent propagator are applied, both qubits are measured. From the observed counts we estimate $p_1(t)=\abs{c_1}^2$ and $p_2(t)=\abs{c_2}^2$. On an ideal circuit, the dynamics would remain within the $\{\ket{10},\ket{01}\}$ subspace. On real hardware, however, noise produces leakage into $\ket{00}$ and $\ket{11}$. To compensate for this effect, we first renormalize the measured population by dividing $p_1(t)$ by $p_1(t)+p_2(t)$.

\begin{figure*}[t]
\centering
\includegraphics[width=0.6\textwidth,height=0.4\textwidth]{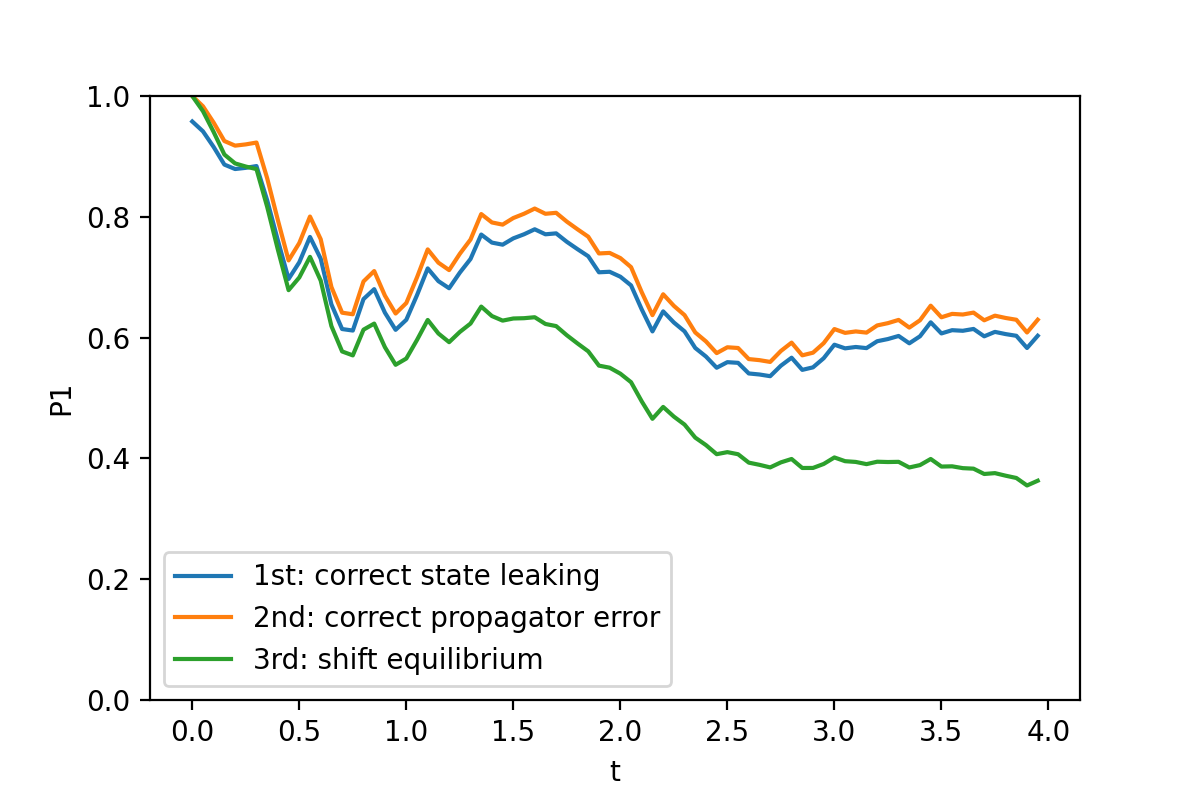}
\caption{Post-processing workflow for the measured population dynamics. The raw normalized population is first corrected so that $p_1(0)=1$, and its long-time limit is then shifted to the finite-temperature equilibrium corresponding to $kT=1$.}
\label{Fig. Post Processing}
\end{figure*}

Furthermore, we correct for state-preparation and short-circuit imperfections by normalizing the trajectory so that $p_1(t=0)=1$. Finally, we account for the fact that the raw noisy-circuit dynamics tend to relax toward an equal-population limit, whereas a biased two-level system at finite temperature should approach a nonequal thermal equilibrium. According to the Boltzmann distribution,
\begin{equation}
Prob(s_i)\propto e^{-E(s_i)/kT}.
\end{equation}
Here, $E(s_i)$ is the energy of state $s_i$, $k$ is the Boltzmann constant, and $T$ is the temperature. Since the two encoded states are biased, the physical equilibrium should satisfy $p_1(\infty)<p_2(\infty)$ when $\ket{s_1}$ is the higher-energy state. In contrast, the raw hardware trajectories typically converge toward $p_1(\infty)=p_2(\infty)=0.5$, corresponding to an effectively infinite-temperature limit. To enforce a finite-temperature equilibrium in post-processing, we apply
\begin{equation}\label{Eq. Correct Equilibrium}
p_{\text{1,fixed}}(t)=e^{-\alpha t}p_1(t)+(1-e^{-\alpha t})Q,
\end{equation}
where $\alpha$ is the decay constant extracted from the population dynamics and $Q=p_1(\infty)$ is the target equilibrium population for the chosen finite-temperature setting. This final transformation shifts the long-time limit of the measured trajectory to the desired equilibrium value. The complete post-processing procedure is summarized in \cref{Fig. Post Processing}.

\section{Benchmark with HEOM}

\subsection{Linear Trotter Steps}
We run our quantum circuit on IBM-Q Jakarta to test our idea. The coefficients of the Hamiltonian used in the quantum simulation are $\epsilon_Q=1.5$ and $J_Q=1$, corresponding to
\begin{equation}
H_Q=\begin{bmatrix} \epsilon_Q&J_Q\\ J_Q&-\epsilon_Q \end{bmatrix}=\begin{bmatrix} 1.5&1\\ 1&-1.5\end{bmatrix}.
\end{equation}
To distinguish the parameters of the quantum simulation from those of the HEOM model, we label quantum-circuit parameters with the subscript $Q$ and HEOM parameters with the subscript $H$.

To determine a suitable Trotter schedule, we first test fixed numbers of Trotter steps, $M=3,5,10$, in the case $\delta_Q=0$, where no noisy identity gates are applied. We compare both simulator and real-device results with the exact dynamics.

\begin{figure*}[t]
\centering
\includegraphics[width=0.95\textwidth,height=0.25\textwidth]{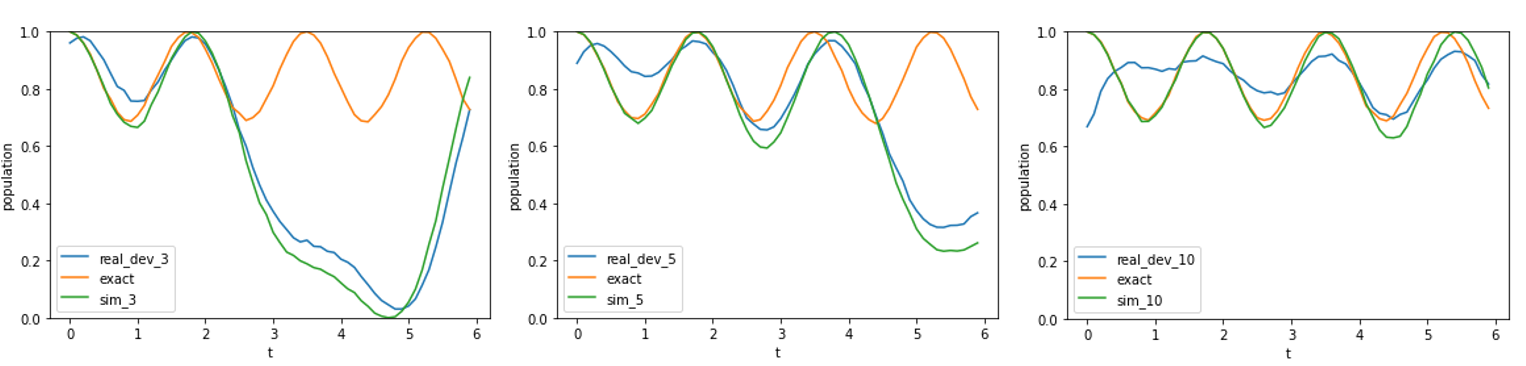}
\caption{Comparison of fixed Trotter schedules with $M=3$, $5$, and $10$. Increasing $M$ reduces Trotter error but increases the circuit depth and hence the accumulated hardware noise.}
\label{Fig. Trotter Step Decision}
\end{figure*}

\begin{figure*}[t]
\centering
\includegraphics[width=0.42\textwidth,height=0.25\textwidth]{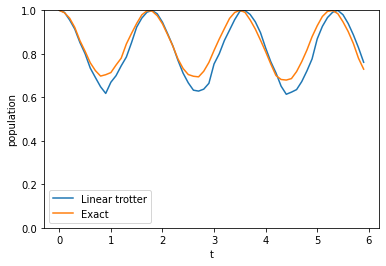}
\caption{Linear Trotter schedule $M=\lceil t/0.4 \rceil$, chosen to keep $\Delta t=t/M$ sufficiently small while avoiding unnecessarily deep circuits.}
\label{Fig. Linear Trotter Simulation}
\end{figure*}

Figure \cref{Fig. Trotter Step Decision} illustrates the tradeoff between Trotter error and hardware noise. When $M=3$, the circuit is shallow, but the time slices are too large and the approximation breaks down at later times. Increasing the number of steps improves the simulator result, but on real hardware the deeper circuit becomes more vulnerable to accumulated gate errors. In particular, the real-device trajectories deteriorate significantly once the circuit depth becomes too large, even in regimes where the simulator remains accurate. Based on this comparison, we adopt a linear Trotter schedule, $M=\lceil t/0.4 \rceil$, which maintains a sufficiently small time step while avoiding unnecessarily deep circuits. As shown in \cref{Fig. Linear Trotter Simulation}, this choice provides a practical compromise for the short-time simulations considered in this work.

\subsection{Comparison with HEOM}
To assess whether the measured quantum-hardware trajectories correspond to meaningful open-system dynamics, we fit them against HEOM trajectories. For each value of $\delta_Q$, we adjust two HEOM parameters, $(\lambda_H,J_H)$, where $\lambda_H$ is the reorganization energy and $J_H$ is the effective tunneling amplitude in the HEOM model. The remaining parameters are fixed throughout the fitting procedure, including $\epsilon_Q=1.5$, $J_Q=1$, $kT=1$, $\epsilon_H=1.5$, and $\gamma_H=11$.

\begin{figure*}[t]
\centering
\subfigure[$\delta_Q=160$]{\includegraphics[width=6cm]{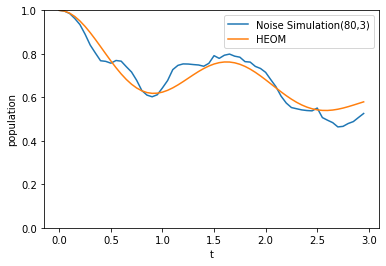}}
\subfigure[$\delta_Q=280$]{\includegraphics[width=6cm]{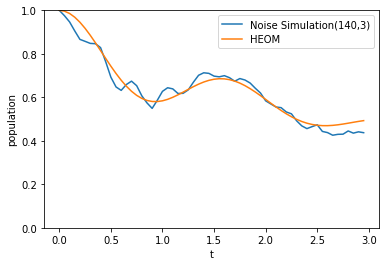}}
\subfigure[$\delta_Q=360$]{\includegraphics[width=6cm]{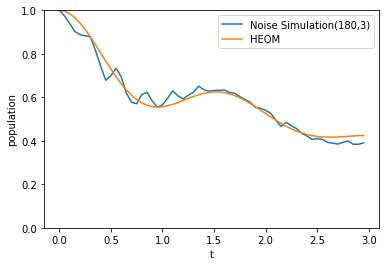}}
\subfigure[$\delta_Q=520$]{\includegraphics[width=6cm]{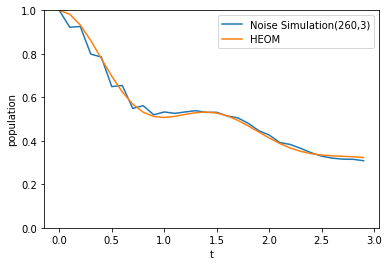}}
\caption{Representative fits between quantum-hardware trajectories and HEOM results for different noisy-gate frequencies: (a) $\delta_Q=160$, (b) $\delta_Q=280$, (c) $\delta_Q=360$, and (d) $\delta_Q=520$. Good agreement is observed across a wide range of effective damping regimes.}
\label{Fig.Fitting Curve (Intermediate)}
\end{figure*}

The representative fits shown in \cref{Fig.Fitting Curve (Intermediate)} indicate that the short-time dynamics generated on the quantum device can be matched well by HEOM trajectories across a range of damping regimes. This agreement does not imply a microscopic equivalence between the hardware noise and the HEOM bath model. Rather, it shows that, after calibration, the noisy two-qubit circuits reproduce population dynamics consistent with an effective dissipative dimer model even in the biased case $\epsilon>0$.

Across the explored values of $\delta_Q$, we find that each quantum-generated trajectory can be associated with a nearby HEOM trajectory, ranging from underdamped to overdamped behavior. Additional fitting examples are provided in the supplementary information.

\subsection{Linear Relation of the Fitting Parameters}
\begin{figure*}[t]
\centering
\includegraphics[width=0.45\textwidth,height=0.3\textwidth]{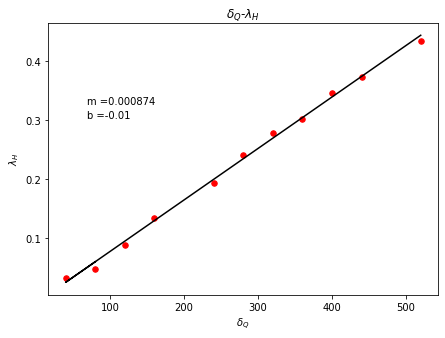}
\includegraphics[width=0.45\textwidth,height=0.3\textwidth]{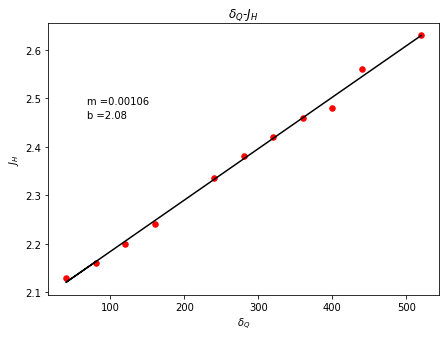}
\caption{Approximately linear relation between the noisy-gate frequency $\delta_Q$ and the fitted HEOM parameters. Here $\delta_Q$ controls the frequency of the identity-gate sequence, while $\lambda_H$ denotes the effective system-bath coupling strength in the Drude-Lorentz model.}
\label{Fig. Linear Relation}
\end{figure*}

When the fitted HEOM parameters are plotted against the noisy-gate frequency, both $\lambda_H$ and $J_H$ display an approximately linear dependence on $\delta_Q$ over the tested regime, as shown in \cref{Fig. Linear Relation}. Empirically, this is important because it suggests that the hardware-noise strength can be calibrated by a simple low-dimensional map onto an effective HEOM description.

We interpret this linearity as a phenomenological relation rather than a proof that the device noise is identical to a Drude-Lorentz bath. Nevertheless, the observed trend indicates that the noisy identity sequence provides a controllable knob for generating families of dissipative trajectories, and that the resulting dynamics can be related in a simple way to the benchmark model.

\subsection{Replacing the HEOM}
The approximate linearity also suggests a stronger and more practical interpretation. We do not claim that the noisy circuit replaces HEOM universally across arbitrary models or parameter regimes. However, within the calibrated family of biased-dimer systems studied here, and under the condition that the relation between $\delta_Q$ and the fitted HEOM parameters remains stable, the noisy two-qubit circuit can serve as an effective surrogate for HEOM in the intermediate regime. Concretely, one may calibrate the map using a finite set of HEOM calculations in representative underdamped and overdamped regimes, and then use interpolation in $\delta_Q$ to generate the intermediate dissipative dynamics without carrying out a new HEOM optimization at every target point. Under these assumptions, the quantum-hardware procedure does more than merely assist HEOM: for this restricted class of problems, it provides a practical replacement strategy for repeated HEOM fitting in the intermediate coupling regime.

\section{Long-term Dynamics Simulation through Transfer Tensor Method}
Although the short-time simulations agree well with HEOM after calibration, long-time simulation remains difficult on current hardware. As the evolution time increases, the number of Trotter steps $M=\lceil t/0.4 \rceil$ also increases, and the resulting deeper circuits accumulate more errors. To address this limitation, we introduce the transfer tensor method (TTM), which uses a short-time training window to infer a memory kernel and extend the dynamics to later times.

\subsection{Transfer Tensor Method}
TTM provides a compact way to extend non-Markovian dynamics from a finite set of short-time data. Once the relevant transfer tensors are inferred from the training trajectories, they can be iterated together with the known short-time states to predict the later-time dynamics.

The set of dynamical maps $\mathcal{E}_k$, where $k=1,\ldots,n_k$, encapsulates the short-time information of a quantum dynamical system~\cite{cerrillo2014non}. For an open quantum system, if we know its density matrix $\rho(t)$ on a short, equally spaced time grid $t_k=k\delta t$, then we can define dynamical maps $\mathcal{E}_k$ through
\begin{equation}\label{generate DM}
\rho(t_k)=\mathcal{E}_k\rho(0).
\end{equation}
Following Ref.~\cite{cerrillo2014non}, we compute transfer tensors $T_n$ that satisfy
\begin{equation}\label{Truncate of Transfer Tensor}
T_n=\mathcal{E}_n-\sum_{m=1}^{n-1}T_{n-m}\mathcal{E}_m,
\end{equation}
for $n=1,\ldots,n_k$. The density matrix can then be propagated according to
\begin{equation}\label{combination}
\rho(t_n)=\sum_{k=0}^{n-1}T_{n-k}\rho(t_k),
\end{equation}
for $n=1,\ldots,n_k$. For integers $n>n_k$, we further assume
\begin{equation}\label{Extending Dynamics}
\rho(t_n)=\sum_{k=0}^{n_k}T_{k'}\rho(t_{n-k'}).
\end{equation}
This truncation is expected to be valid when the underlying memory kernel decays sufficiently quickly. Under that assumption, the information contained in the training window is enough to predict the subsequent states, allowing the short-time data to be propagated to later times.

\subsection{Extending the Result from Quantum Computer}
In our setting, TTM is used to extend the short-time quantum trajectories beyond the range where direct circuit simulation remains reliable. The goal is not to eliminate all hardware errors, but to determine whether the information contained in the accurately measured early-time dynamics is sufficient to reconstruct the later-time evolution.

\subsubsection{Generating Dynamical Map Set}
In our system, the density matrix is $2\times 2$. After vectorization, each dynamical map $\mathcal{E}_k$ becomes a $4\times 4$ matrix. Reconstructing these maps requires four linearly independent sets of density matrices, which we generate from the four initial states
\begin{equation}\label{Four Init State}
\rho_S(t=0)=\frac{1}{2}(I+Z),\ \frac{1}{2}(I-Z),\ \frac{1}{2}(I+X),\ \frac{1}{2}(I+Y).
\end{equation}
Without full state tomography, the off-diagonal terms of $\rho_S(t)$ are not directly accessible from population measurements alone. To construct an approximate dynamical map, we therefore infer these off-diagonal elements through a model-based reconstruction. Since the density matrix is expressed in the site basis, we first transform it to the eigenbasis $\rho_E(t)$ of the fixed Hamiltonian:
\begin{equation}\label{basis_transform}
U^{\dagger}\rho_S(t)U=\rho_E(t),
\end{equation}
where $U$ diagonalizes the system Hamiltonian. For $\rho_E(t)$, we assume
\begin{equation}\label{Infer Diagonal}
\alpha=\sqrt{p_1\cdot p_2}e^{i\omega t}e^{-\alpha t},
\end{equation}
where $\alpha$ and $\omega$, representing the decay constant and oscillation frequency of $P_1$, are obtained from fits to the site-basis population dynamics. Combining Eqs.~(\ref{basis_transform}) and (\ref{Infer Diagonal}) gives an approximate reconstruction of the off-diagonal elements of $\rho_S(t)$, which in turn allows us to solve for the dynamical maps through \cref{generate DM}. Once the dynamical maps are obtained, the transfer tensors can be constructed and used in \cref{Extending Dynamics} to extend the dynamics.

\section{Validity of Transfer Tensor Method}
We test TTM on both simulator data and data obtained from the IBMQ Jakarta device. In each case, the method uses the four initial-state trajectories listed in \cref{Four Init State}. Because the experiment measures only the populations of $\ket{10}$ and $\ket{01}$, the off-diagonal elements needed for TTM are supplied by the reconstruction procedure based on Eqs.~(\ref{basis_transform}) and (\ref{Infer Diagonal}).

\subsection{Quantum Simulator}
\begin{figure*}[t]
\centering
\includegraphics[width=0.6\textwidth]{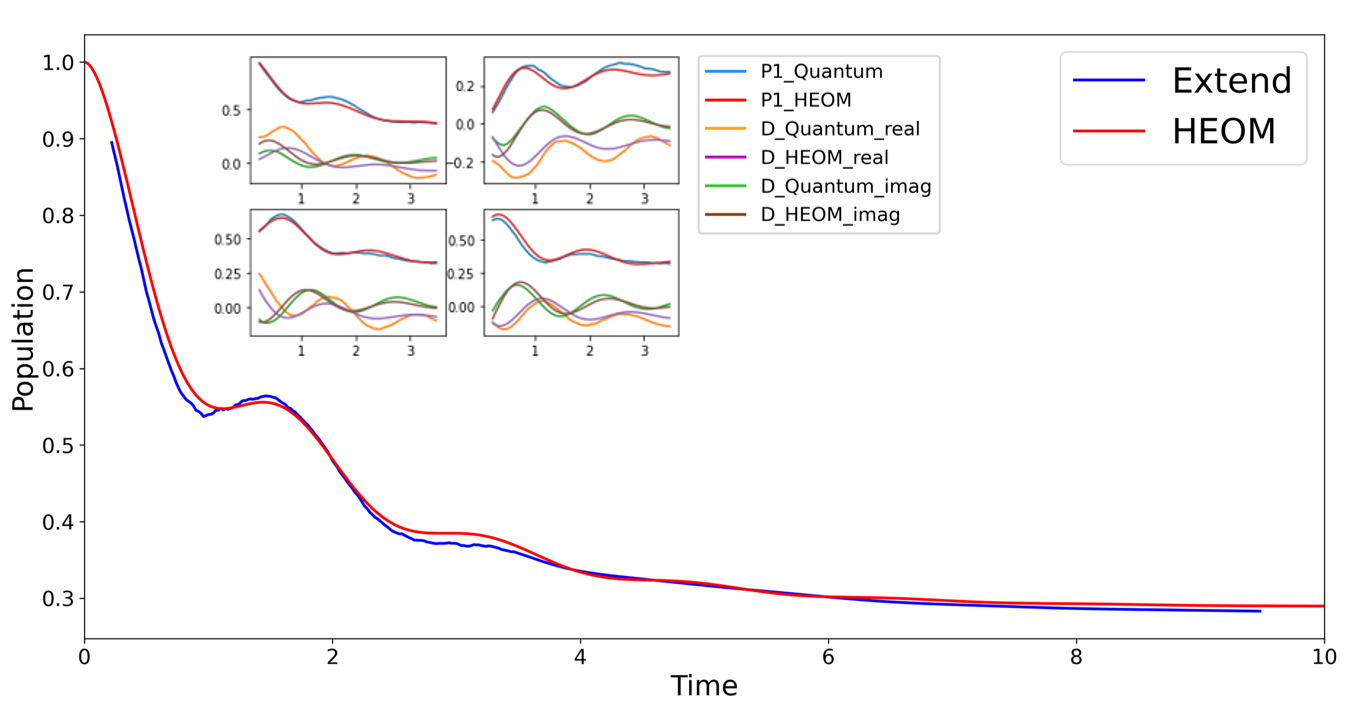}
\caption{TTM extension for simulator data at $\delta_Q=200$. The inset shows the short-time training data, while the main panel compares the TTM-extended trajectory with the corresponding HEOM result. The agreement remains good well beyond the training window.}
\label{Fig.TTM-Simulating data}
\end{figure*}

We first test the method using Qiskit simulator data with depolarizing error rate 0.002 and $\delta_Q=200$, focusing on the initial state $\rho=0.5(I+Z)$. In \cref{Fig.TTM-Simulating data}, the short-time training trajectories agree closely with HEOM over $0.1\leq t\leq 3.5$, and the TTM extension continues to track the HEOM result well up to $t=9$. This corresponds to an extension to nearly three times the duration of the original training window.

\begin{figure*}[t]
\centering
\subfigure[$\delta_Q=150$]{\includegraphics[width=0.45\textwidth]{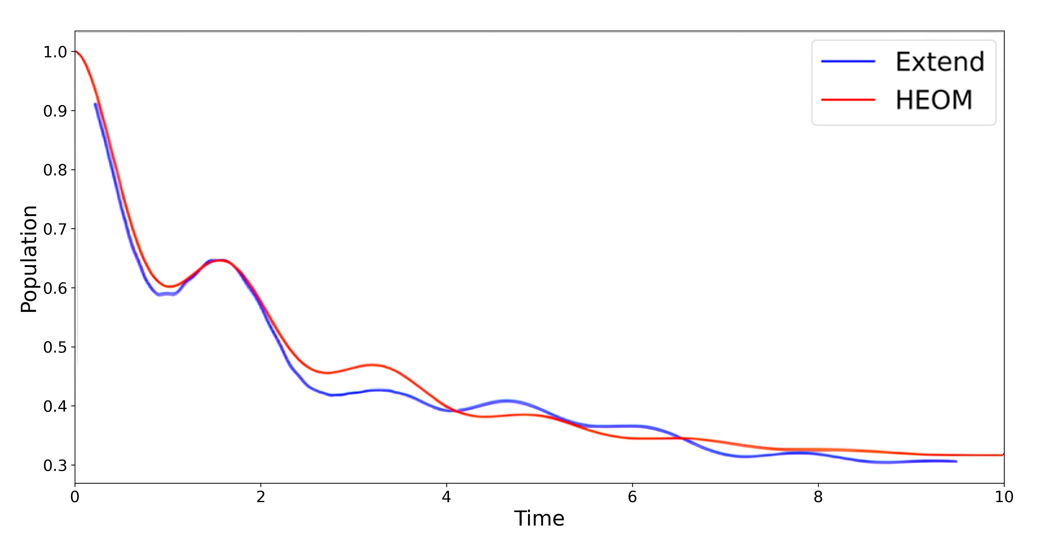}}
\subfigure[$\delta_Q=200$]{\includegraphics[width=0.45\textwidth]{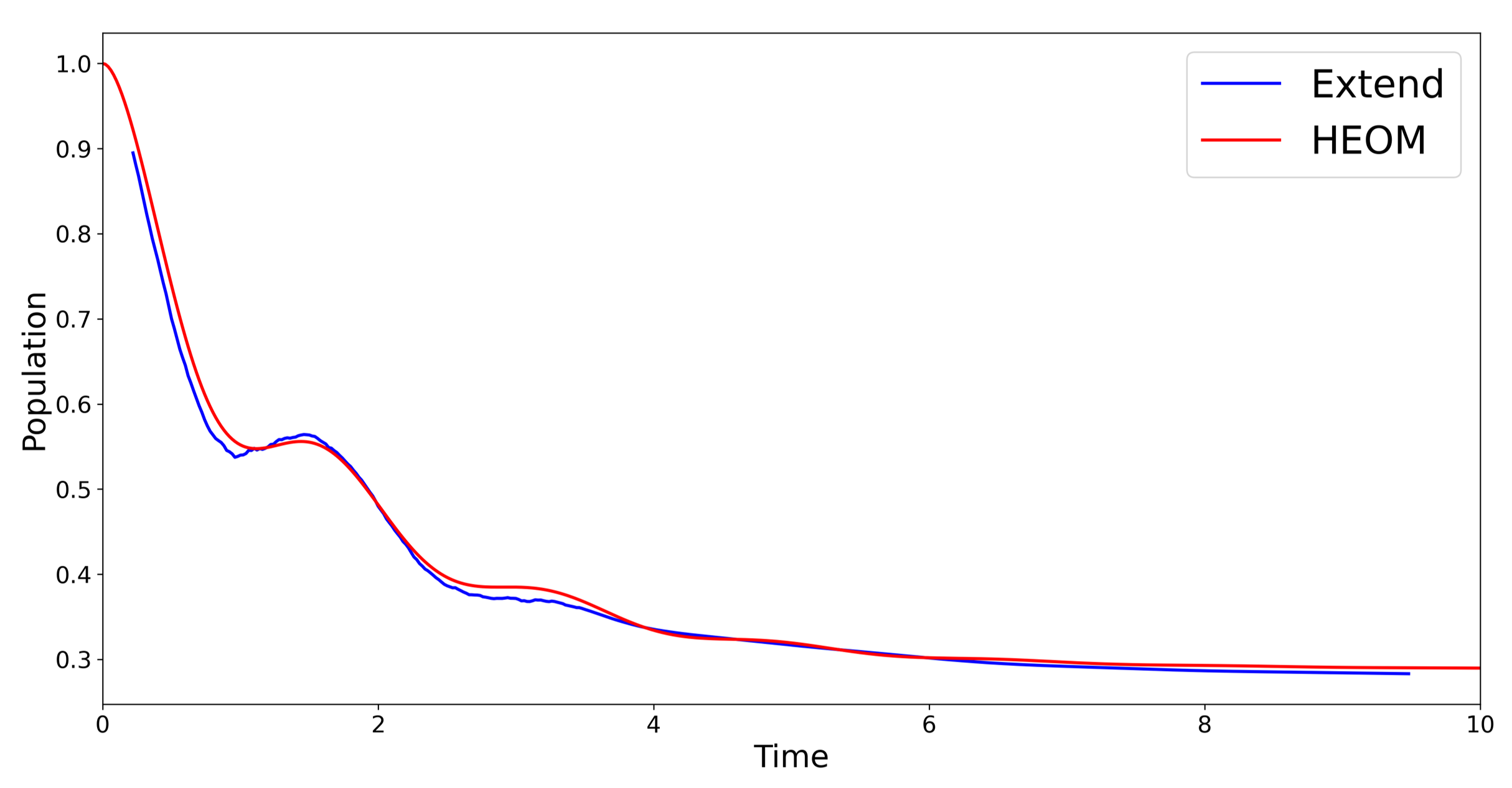}}
\subfigure[$\delta_Q=250$]{\includegraphics[width=0.45\textwidth]{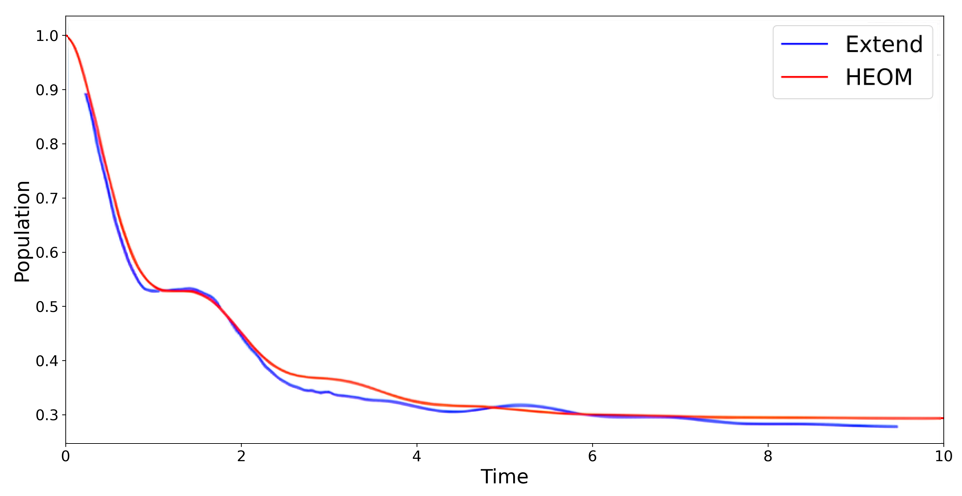}}
\subfigure[$\delta_Q=300$]{\includegraphics[width=0.45\textwidth]{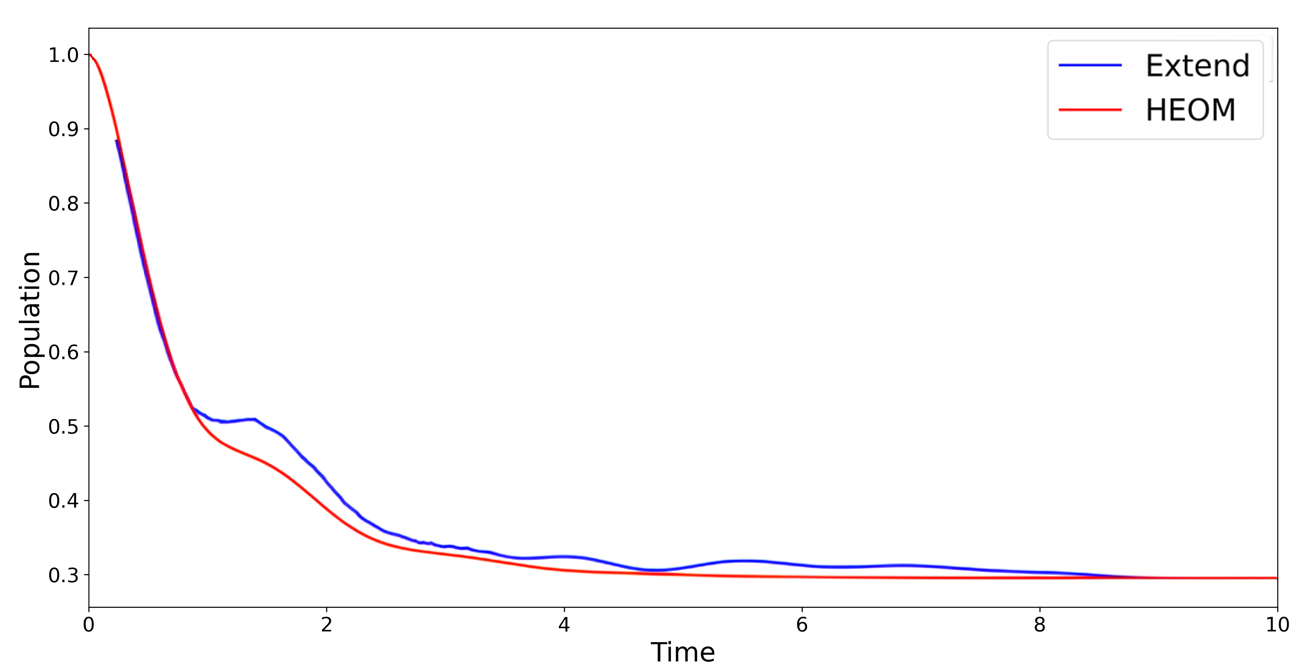}}
\caption{Additional simulator examples comparing TTM-extended trajectories with HEOM. In most tested cases, the extension remains accurate over times substantially longer than the original training window.}
\label{Fig.TTM-Simulating data others}
\end{figure*}

\subsection{Real Device}
The real-device results are more challenging. Although the TTM procedure works well on simulator data, it is difficult to reproduce the same level of agreement on IBMQ Jakarta once $\delta_Q>0$. The main difficulty comes from the trajectories initialized in $\rho_3=0.5(I+X)$ and $\rho_4=0.5(I+Y)$, which are coherence-sensitive superposition states. These trajectories are substantially more fragile to hardware noise than those initialized in the site-basis states $\ket{01}$ and $\ket{10}$, and they therefore degrade the quality of the reconstructed dynamical maps. Consistent with this interpretation, the four initial-state trajectories agree best with HEOM in the limiting case $\delta_Q=0$, where the dissipative component is absent.

\begin{figure*}[t]
\centering
\includegraphics[width=0.8\textwidth]{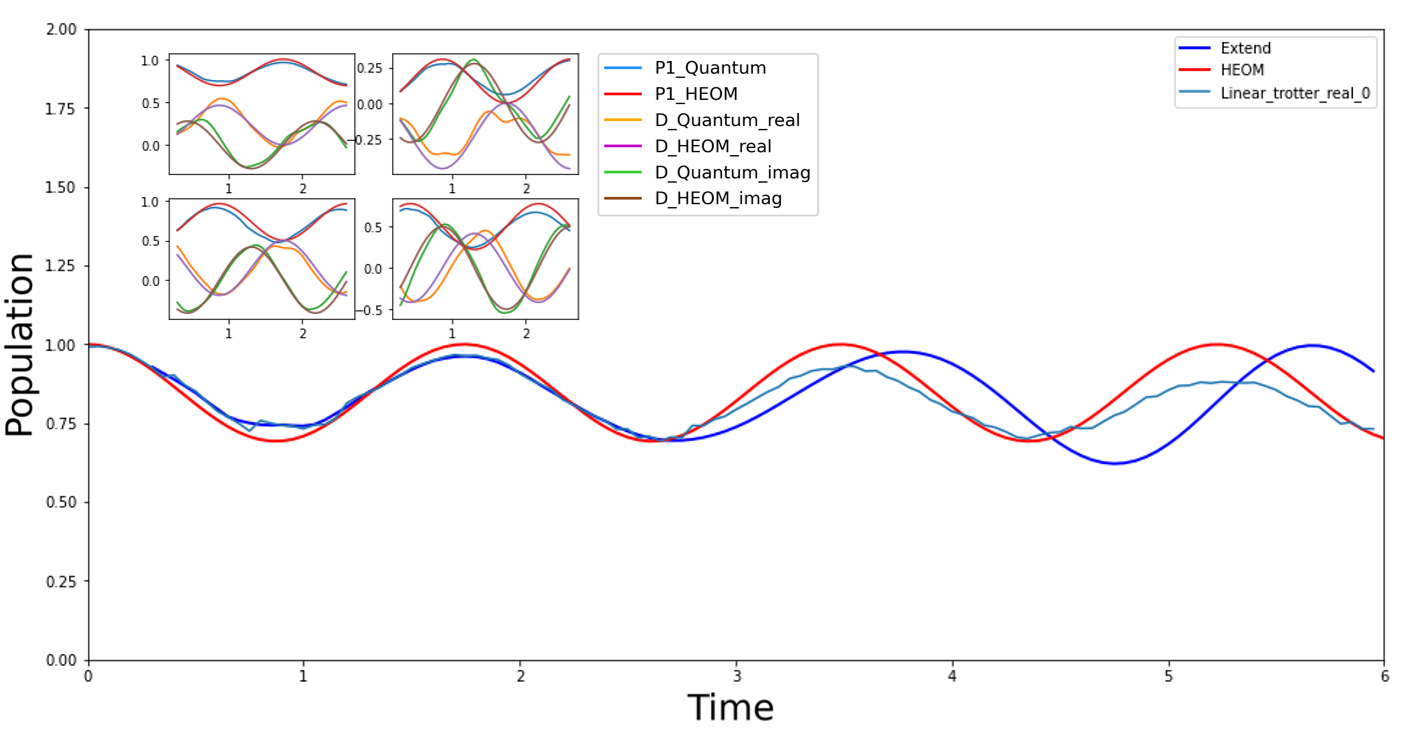}
\caption{Comparison of HEOM, direct real-device simulation, and the TTM extension constructed from short-time real-device data. The TTM result preserves amplitude better than direct long-time execution, but its frequency remains inaccurate.}
\label{Fig.TTM-Real perform}
\end{figure*}

To assess whether TTM can nevertheless act as a partial error-mitigation strategy, we compare the following three dynamics:
\begin{itemize}
\item dynamics extended to $0<t<6$ generated by TTM from real-device data over $0<t<2.5$,
\item direct real-device dynamics over $0<t<6$,
\item HEOM dynamics over $0<t<6$.
\end{itemize}
Using the short-time window $0\leq t\leq 2.5$ as training data, we extend the dynamics to $0<t<6$, as shown in \cref{Fig.TTM-Real perform}. Relative to the directly measured real-device dynamics, the TTM-extended trajectories preserve the oscillation amplitude more effectively, which indicates that the early-time data retain useful information about the underlying dynamics. At the same time, the extended trajectories do not reproduce the correct oscillation frequency with sufficient accuracy, likely because the training window is too short and the coherence-sensitive input data are already significantly degraded. Therefore, on real hardware TTM does not outperform direct simulation in the present implementation, but it does indicate a plausible direction for extending noisy short-time quantum data.

\begin{figure*}[t]
\centering
\includegraphics[width=0.6\textwidth]{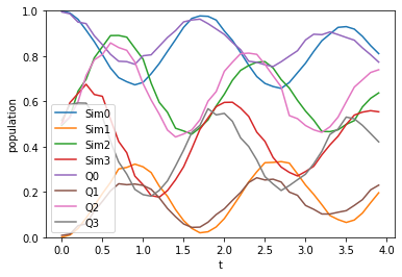}
\caption{Real-device trajectories for the four initial density matrices $\rho_1=(I+Z)/2$, $\rho_2=(I-Z)/2$, $\rho_3=(I+X)/2$, and $\rho_4=(I+Y)/2$. The coherence-sensitive states $\rho_3$ and $\rho_4$ become substantially noisier once $\delta_Q>0$, limiting the quality of the TTM reconstruction on hardware.}
\label{Fig.TTM-Real data}
\end{figure*}

\section{Concluding Remarks}
This work shows that dissipative energy-transfer dynamics in a biased exciton dimer can be simulated on present-day noisy quantum hardware in a controlled and interpretable way when device noise is treated as a calibrated effective resource. Using only two qubits on IBMQ Jakarta, we reproduce short-time dissipative trajectories that can be benchmarked quantitatively against HEOM over the tested parameter regime.

The central result is not merely that hardware noise can mimic damping qualitatively, but that the noisy-circuit output admits a simple empirical calibration to an effective HEOM description. Over the biased-dimer family studied here, the fitted HEOM parameters vary approximately linearly with the noisy-gate frequency. Under the explicit condition that this calibration remains stable within the target family of systems, the quantum-hardware procedure can do more than assist classical modeling: it can serve as a practical surrogate for repeated HEOM fitting at intermediate parameter points after a finite calibration set has been established. This is the sense in which the present approach can replace HEOM in part, namely as a calibrated interpolation tool within a restricted but well-defined regime rather than as a universal substitute for numerically exact open-system solvers.

For long-time dynamics, the transfer tensor method provides a natural way to leverage short-time quantum data beyond the circuit-depth limit. In simulator studies, TTM extends the dynamics substantially while maintaining good agreement with HEOM, demonstrating that the short-time quantum trajectories contain enough information for accurate non-Markovian extrapolation. On real hardware, the main limitation is not the TTM framework itself but the quality of the coherence-sensitive initial states required for reconstructing the dynamical maps. Even with that limitation, the TTM-extended real-device trajectories retain physically meaningful features more successfully than direct long-time execution.

Taken together, these results support a stronger view of NISQ-era open-system simulation: noisy few-qubit devices need not be useful only after full error suppression, but can already function as calibrated phenomenological simulators when combined with reliable benchmarking and data-driven long-time extensions. For the biased-dimer problem examined here, that strategy yields not only qualitative emulation of dissipation, but a practically actionable route toward replacing a portion of the classical HEOM workload in the intermediate regime.

\nocite{*}
\bibliographystyle{apsrev4-2}
\bibliography{references}

@article{cerrillo2014non,
  title={Non-Markovian dynamical maps: numerical processing of open quantum trajectories},
  author={Cerrillo, Javier and Cao, Jianshu},
  journal={Physical review letters},
  volume={112},
  number={11},
  pages={110401},
  year={2014},
  publisher={APS}
}

@misc{sun2021efficient,
      title={Efficient Quantum Simulation of Open Quantum System Dynamics on Noisy Quantum Computers}, 
      author={Shin Sun and Li-Chai Shih and Yuan-Chung Cheng},
      year={2021},
      eprint={2106.12882},
      archivePrefix={arXiv},
      primaryClass={quant-ph}
}

@article{nielsen2001quantum,
  title={Quantum computation and quantum information},
  author={Nielsen, Michael A and Chuang, Isaac L},
  journal={Phys. Today},
  volume={54},
  number={2},
  pages={60},
  year={2001}
}

@article{bacon2001universal,
  title={Universal simulation of Markovian quantum dynamics},
  author={Bacon, Dave and Childs, Andrew M and Chuang, Isaac L and Kempe, Julia and Leung, Debbie W and Zhou, Xinlan},
  journal={Physical Review A},
  volume={64},
  number={6},
  pages={062302},
  year={2001},
  publisher={APS}
}

@article{georgescu2014quantum,
  title={Quantum simulation},
  author={Georgescu, Iulia M and Ashhab, Sahel and Nori, Franco},
  journal={Reviews of Modern Physics},
  volume={86},
  number={1},
  pages={153},
  year={2014},
  publisher={APS}
}

@article{altman2021quantum,
  title={Quantum simulators: Architectures and opportunities},
  author={Altman, Ehud and Brown, Kenneth R and Carleo, Giuseppe and Carr, Lincoln D and Demler, Eugene and Chin, Cheng and DeMarco, Brian and Economou, Sophia E and Eriksson, Mark A and Fu, Kai-Mei C and others},
  journal={PRX Quantum},
  volume={2},
  number={1},
  pages={017003},
  year={2021},
  publisher={APS}
}

@article{yung2014introduction,
  title={Introduction to quantum algorithms for physics and chemistry},
  author={Yung, Man-Hong and Whitfield, James D and Boixo, Sergio and Tempel, David G and Aspuru-Guzik, Al{\'a}n},
  journal={Quantum Information and Computation for Chemistry},
  pages={67--106},
  year={2014},
  publisher={Wiley Online Library}
}

@article{o2016scalable,
  title={Scalable quantum simulation of molecular energies},
  author={O’Malley, Peter JJ and Babbush, Ryan and Kivlichan, Ian D and Romero, Jonathan and McClean, Jarrod R and Barends, Rami and Kelly, Julian and Roushan, Pedram and Tranter, Andrew and Ding, Nan and others},
  journal={Physical Review X},
  volume={6},
  number={3},
  pages={031007},
  year={2016},
  publisher={APS}
}

@article{colless2018computation,
  title={Computation of molecular spectra on a quantum processor with an error-resilient algorithm},
  author={Colless, James I and Ramasesh, Vinay V and Dahlen, Dar and Blok, Machiel S and Kimchi-Schwartz, Mollie E and McClean, Jarrod R and Carter, Jonathan and de Jong, Wibe A and Siddiqi, Irfan},
  journal={Physical Review X},
  volume={8},
  number={1},
  pages={011021},
  year={2018},
  publisher={APS}
}

@article{cao2019quantum,
  title={Quantum chemistry in the age of quantum computing},
  author={Cao, Yudong and Romero, Jonathan and Olson, Jonathan P and Degroote, Matthias and Johnson, Peter D and Kieferov{\'a}, M{\'a}ria and Kivlichan, Ian D and Menke, Tim and Peropadre, Borja and Sawaya, Nicolas PD and others},
  journal={Chemical reviews},
  volume={119},
  number={19},
  pages={10856--10915},
  year={2019},
  publisher={ACS Publications}
}

@article{mcardle2020quantum,
  title={Quantum computational chemistry},
  author={McArdle, Sam and Endo, Suguru and Aspuru-Guzik, Al{\'a}n and Benjamin, Simon C and Yuan, Xiao},
  journal={Reviews of Modern Physics},
  volume={92},
  number={1},
  pages={015003},
  year={2020},
  publisher={APS}
}

@article{preskill2018quantum,
  title={Quantum computing in the NISQ era and beyond},
  author={Preskill, John},
  journal={Quantum},
  volume={2},
  pages={79},
  year={2018},
  publisher={Verein zur F{\"o}rderung des Open Access Publizierens in den Quantenwissenschaften}
}

@inproceedings{murali2019noise,
  title={Noise-adaptive compiler mappings for noisy intermediate-scale quantum computers},
  author={Murali, Prakash and Baker, Jonathan M and Javadi-Abhari, Ali and Chong, Frederic T and Martonosi, Margaret},
  booktitle={Proceedings of the twenty-fourth international conference on architectural support for programming languages and operating systems},
  pages={1015--1029},
  year={2019}
}

@article{knill1998resilient,
  title={Resilient quantum computation},
  author={Knill, Emanuel and Laflamme, Raymond and Zurek, Wojciech H},
  journal={Science},
  volume={279},
  number={5349},
  pages={342--345},
  year={1998},
  publisher={American Association for the Advancement of Science}
}

@article{bravyi2018correcting,
  title={Correcting coherent errors with surface codes},
  author={Bravyi, Sergey and Englbrecht, Matthias and K{\"o}nig, Robert and Peard, Nolan},
  journal={npj Quantum Information},
  volume={4},
  number={1},
  pages={55},
  year={2018},
  publisher={Nature Publishing Group UK London}
}

@book{breuer2002theory,
  title={The theory of open quantum systems},
  author={Breuer, Heinz-Peter and Petruccione, Francesco},
  year={2002},
  publisher={Oxford University Press, USA}
}

@book{weiss2012quantum,
  title={Quantum dissipative systems},
  author={Weiss, Ulrich},
  year={2012},
  publisher={World Scientific}
}

@article{croce2020light,
  title={Light harvesting in oxygenic photosynthesis: Structural biology meets spectroscopy},
  author={Croce, Roberta and van Amerongen, Herbert},
  journal={Science},
  volume={369},
  number={6506},
  pages={eaay2058},
  year={2020},
  publisher={American Association for the Advancement of Science}
}

@article{arsenault2020vibronic,
  title={Vibronic mixing enables ultrafast energy flow in light-harvesting complex II},
  author={Arsenault, Eric A and Yoneda, Yusuke and Iwai, Masakazu and Niyogi, Krishna K and Fleming, Graham R},
  journal={Nature communications},
  volume={11},
  number={1},
  pages={1460},
  year={2020},
  publisher={Nature Publishing Group UK London}
}

@article{wang2019quantum,
  title={Quantum coherences reveal excited-state dynamics in biophysical systems},
  author={Wang, Lili and Allodi, Marco A and Engel, Gregory S},
  journal={Nature Reviews Chemistry},
  volume={3},
  number={8},
  pages={477--490},
  year={2019},
  publisher={Nature Publishing Group UK London}
}

@article{rafiq2018fundamental,
  title={From fundamental theories to quantum coherences in electron transfer},
  author={Rafiq, Shahnawaz and Scholes, Gregory D},
  journal={Journal of the American Chemical Society},
  volume={141},
  number={2},
  pages={708--722},
  year={2018},
  publisher={ACS Publications}
}

@article{kim2019ultrafast,
  title={Ultrafast charge transfer coupled with lattice phonons in two-dimensional covalent organic frameworks},
  author={Kim, Tae Wu and Jun, Sunhong and Ha, Yoonhoo and Yadav, Rajesh K and Kumar, Abhishek and Yoo, Chung-Yul and Oh, Inhwan and Lim, Hyung-Kyu and Shin, Jae Won and Ryoo, Ryong and others},
  journal={Nature Communications},
  volume={10},
  number={1},
  pages={1873},
  year={2019},
  publisher={Nature Publishing Group UK London}
}

@article{zurek2003decoherence,
  title={Decoherence, einselection, and the quantum origins of the classical},
  author={Zurek, Wojciech Hubert},
  journal={Reviews of modern physics},
  volume={75},
  number={3},
  pages={715},
  year={2003},
  publisher={APS}
}

@article{tanimura1989time,
  title={Time evolution of a quantum system in contact with a nearly Gaussian-Markoffian noise bath},
  author={Tanimura, Yoshitaka and Kubo, Ryogo},
  journal={Journal of the Physical Society of Japan},
  volume={58},
  number={1},
  pages={101--114},
  year={1989},
  publisher={The Physical Society of Japan}
}

@article{tanimura2006stochastic,
  title={Stochastic Liouville, Langevin, Fokker--Planck, and master equation approaches to quantum dissipative systems},
  author={Tanimura, Yoshitaka},
  journal={Journal of the Physical Society of Japan},
  volume={75},
  number={8},
  pages={082001},
  year={2006},
  publisher={The Physical Society of Japan}
}

@article{ishizaki2009unified,
  title={Unified treatment of quantum coherent and incoherent hopping dynamics in electronic energy transfer: Reduced hierarchy equation approach},
  author={Ishizaki, Akihito and Fleming, Graham R},
  journal={The Journal of chemical physics},
  volume={130},
  number={23},
  year={2009},
  publisher={AIP Publishing}
}

@article{jin2008exact,
  title={Exact dynamics of dissipative electronic systems and quantum transport: Hierarchical equations of motion approach},
  author={Jin, Jinshuang and Zheng, Xiao and Yan, YiJing},
  journal={The Journal of chemical physics},
  volume={128},
  number={23},
  year={2008},
  publisher={AIP Publishing}
}

@article{maniscalco2004simulating,
  title={Simulating quantum Brownian motion with single trapped ions},
  author={Maniscalco, Sabrina and Piilo, Jyrki and Intravaia, F and Petruccione, F and Messina, A},
  journal={Physical Review A},
  volume={69},
  number={5},
  pages={052101},
  year={2004},
  publisher={APS}
}

@article{chiuri2012linear,
  title={Linear optics simulation of quantum non-Markovian dynamics},
  author={Chiuri, Andrea and Greganti, Chiara and Mazzola, Laura and Paternostro, Mauro and Mataloni, Paolo},
  journal={Scientific reports},
  volume={2},
  number={1},
  pages={968},
  year={2012},
  publisher={Nature Publishing Group UK London}
}

@article{mostame2012quantum,
  title={Quantum simulator of an open quantum system using superconducting qubits: exciton transport in photosynthetic complexes},
  author={Mostame, Sarah and Rebentrost, Patrick and Eisfeld, Alexander and Kerman, Andrew J and Tsomokos, Dimitris I and Aspuru-Guzik, Al{\'a}n},
  journal={New Journal of Physics},
  volume={14},
  number={10},
  pages={105013},
  year={2012},
  publisher={IOP Publishing}
}

@misc{anton2018studying,
  title={Studying light-harvesting models with superconducting circuits Nat},
  author={Anton, P and others},
  year={2018},
  publisher={Commun}
}

@article{wang2018efficient,
  title={Efficient quantum simulation of photosynthetic light harvesting},
  author={Wang, Bi-Xue and Tao, Ming-Jie and Ai, Qing and Xin, Tao and Lambert, Neill and Ruan, Dong and Cheng, Yuan-Chung and Nori, Franco and Deng, Fu-Guo and Long, Gui-Lu},
  journal={NPJ Quantum Information},
  volume={4},
  number={1},
  pages={52},
  year={2018},
  publisher={Nature Publishing Group UK London}
}

@article{trautmann2018trapped,
  title={Trapped-ion quantum simulation of excitation transport: Disordered, noisy, and long-range connected quantum networks},
  author={Trautmann, Nils and Hauke, Philipp},
  journal={Physical Review A},
  volume={97},
  number={2},
  pages={023606},
  year={2018},
  publisher={APS}
}

@article{feynman2018simulating,
  title={Simulating physics with computers},
  author={Feynman, Richard P and others},
  journal={Int. j. Theor. phys},
  volume={21},
  number={6/7},
  year={2018}
}

@article{lloyd1996universal,
  title={Universal quantum simulators},
  author={Lloyd, Seth},
  journal={Science},
  volume={273},
  number={5278},
  pages={1073--1078},
  year={1996},
  publisher={American Association for the Advancement of Science}
}

@article{maier2019environment,
  title={Environment-assisted quantum transport in a 10-qubit network},
  author={Maier, Christine and Brydges, Tiff and Jurcevic, Petar and Trautmann, Nils and Hempel, Cornelius and Lanyon, Ben P and Hauke, Philipp and Blatt, Rainer and Roos, Christian F},
  journal={Physical review letters},
  volume={122},
  number={5},
  pages={050501},
  year={2019},
  publisher={APS}
}

@article{su2020quantum,
  title={Quantum algorithm for the simulation of open-system dynamics and thermalization},
  author={Su, Hong-Yi and Li, Ying},
  journal={Physical Review A},
  volume={101},
  number={1},
  pages={012328},
  year={2020},
  publisher={APS}
}

@article{garcia2020ibm,
  title={IBM Q Experience as a versatile experimental testbed for simulating open quantum systems},
  author={Garc{\'\i}a-P{\'e}rez, Guillermo and Rossi, Matteo AC and Maniscalco, Sabrina},
  journal={npj Quantum Information},
  volume={6},
  number={1},
  pages={1},
  year={2020},
  publisher={Nature Publishing Group UK London}
}

@article{rost2020simulation,
  title={Simulation of thermal relaxation in spin chemistry systems on a quantum computer using inherent qubit decoherence},
  author={Rost, Brian and Jones, Barbara and Vyushkova, Mariya and Ali, Aaila and Cullip, Charlotte and Vyushkov, Alexander and Nabrzyski, Jarek},
  journal={arXiv preprint arXiv:2001.00794},
  year={2020}
}

\appendix

\section{Identity-Gate Error Characterization}
To understand how IBM-Q hardware noise enters our simulation, we further examine the character of the single-qubit gate errors. Prior work showed that these errors are not well described by purely random depolarizing noise. If the errors were purely depolarizing, repeated application of a nominal identity sequence such as $XX=I$ would simply shrink the Bloch vector toward the origin without inducing a coherent rotation. Experimentally, however, repeated application of the gate sequence produces a noticeable rotation, indicating the presence of coherent over-rotation or other non-depolarizing error components. Motivated by this observation, we repeated the same type of test on IBM-Q Jakarta.

\begin{figure}[t]
\centering
\includegraphics[width=0.32\columnwidth]{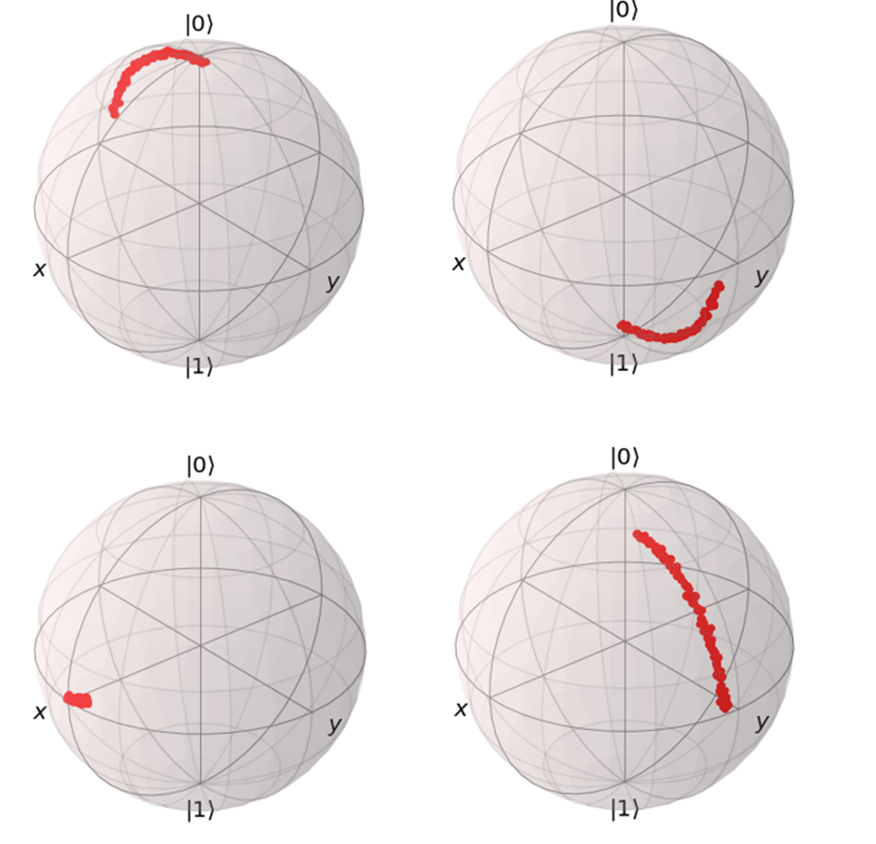}
\includegraphics[width=0.32\columnwidth]{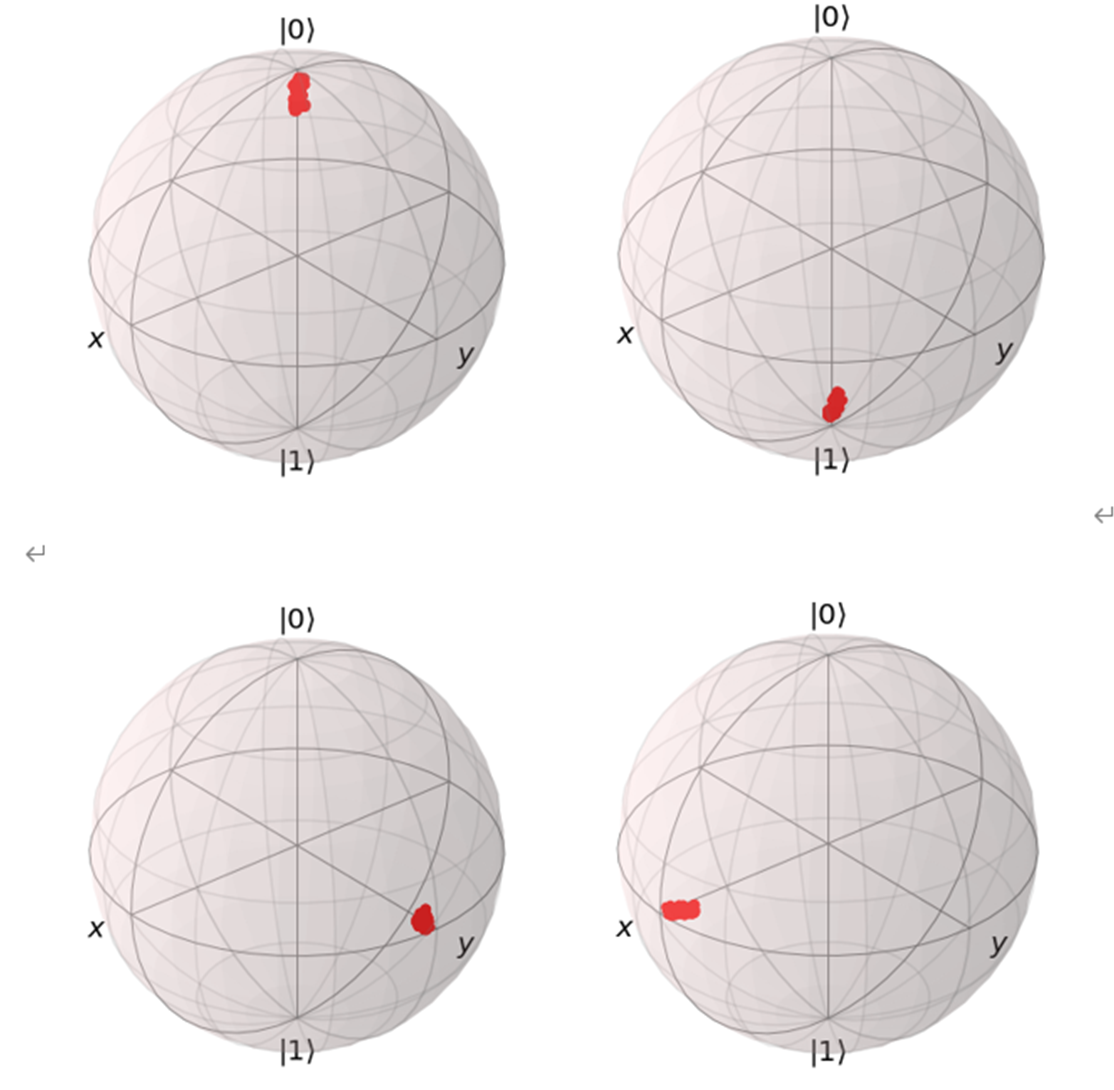}
\caption{Single-qubit error characterization on IBM-Q Jakarta. Left: repeated application of $XX$. Right: repeated application of $XZXZ$. The observed rotations indicate the presence of coherent, non-depolarizing error components.}
\label{Fig.XX and XZXZ}
\end{figure}

By inserting additional $Z$ gates, one can partially compensate the coherent over-rotation induced by the repeated $X$ gates and obtain a noise process that is closer to pure depolarization.

\begin{figure}[b]
\centering
\includegraphics[width=0.42\columnwidth]{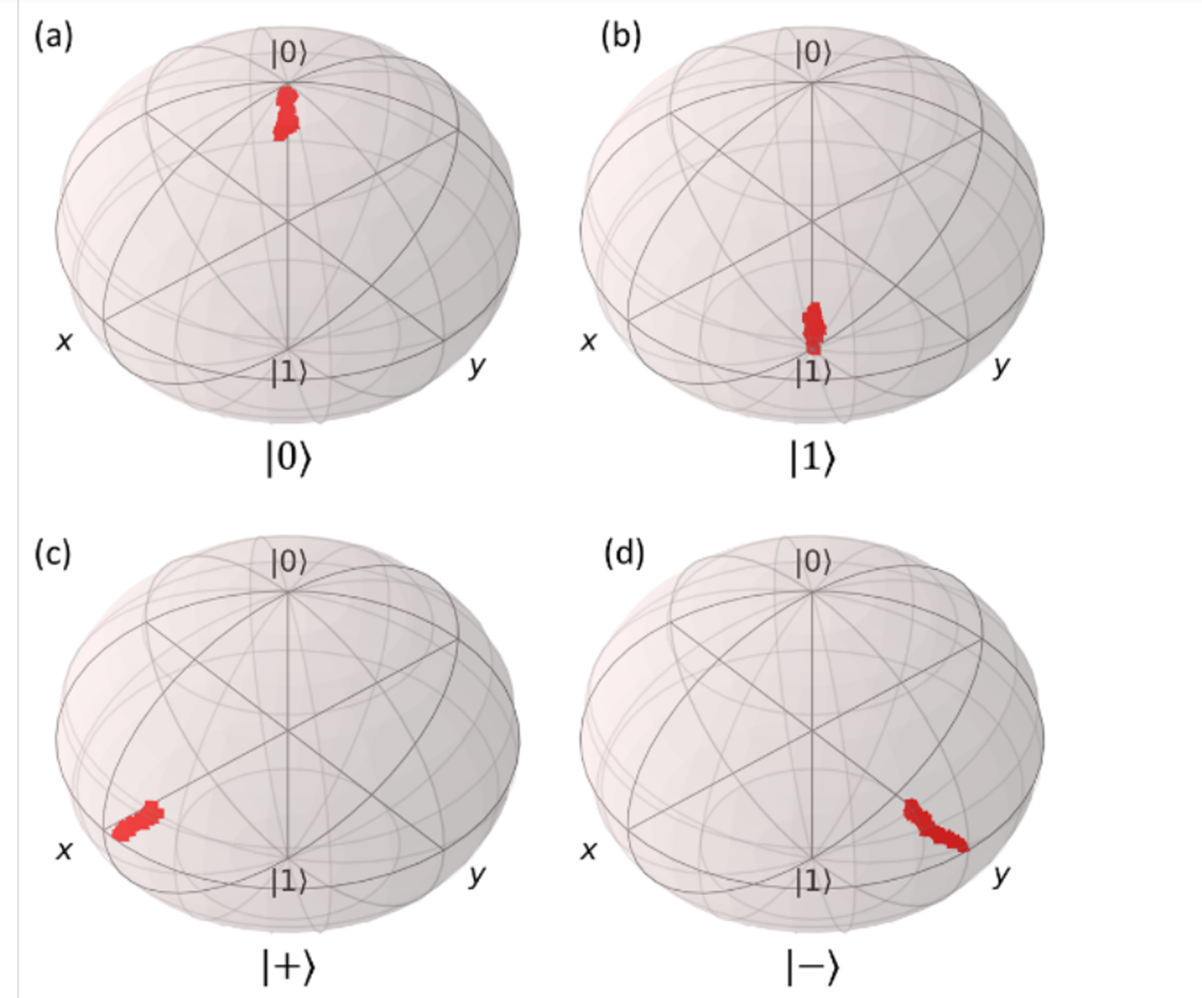}
\caption{Repeated application of the compensated identity sequence $XZXZZXZXZZ$ on a single qubit from different initial states on IBM-Q Jakarta. The added $Z$ gates suppress the coherent rotation seen in the simpler sequences.}
\label{Fig.XZXZZ}
\end{figure}

As indicated by \cref{Fig.XZXZZ}, the sequence $I=(XZXZZ)^2$ provides a controllable noisy identity operation with reduced coherent rotation. This is the gate sequence used in the main text to generate the effective dissipative channel in our open-system simulation.

\end{document}